\documentstyle[prb,aps]{revtex}
\begin{document}

% =================================================================== %  Front matter

\draft

\title{
  Quantum Coherence and Decoherence in Magnetic Nanostructures}

\author{Eugene M. Chudnovsky} 
\address{Department of Physics and Astronomy, City University of
  New York--Lehman College, \\ Bedford Park Boulevard West, Bronx, New
  York 10468-1589 }

\date{\today}

\maketitle

\begin{abstract}
  The prospect of developing magnetic qubits is discussed. The first part
  of the article makes suggestions on how to achieve the coherent quantum superposition 
  of spin states in small ferromagnetic clusters, weakly uncompensated 
  antiferromagnetic clusters, and magnetic molecules. The second part of the 
  article deals with mechanisms of decoherence expected in magnetic systems. 
  Main decohering effects are coming from nuclear spins and magnetic fields. 
  They can be reduced by isotopic purification and superconducting shielding. In
  that case the time reversal symmetry of spin Hamiltonians makes spin-phonon
  coupling ineffective in destroying quantum coherence. 
\end{abstract}


% =================================================================== 
%  Body of the document

\section{Introduction}

Is quantum computing physics or engineering? Some time ago I asked this question 
to Murray Gell-Mann who replied that he had a formal proof that it is physics: 
articles on quantum computing are published by The Physical Review Letters. In the
spirit of his answer I will concentrate on the physics aspects of the problem,
leaving out the engineering aspects. The purpose of this article is to discuss
the prospect of using small magnetic clusters as qubits.

Qubits based upon quantum superposition of the $|\uparrow>$ and $|\downarrow>$ 
states of individual electrons and nuclei have been discussed during all 
years of quantum computermania \cite{Lo}. In parallel, but with little overlap,
there has been intensive theoretical and experimental research on spin tunneling 
between the $|\uparrow>$ and $|\downarrow>$ states of 
molecular magnets and in ferromagnetic and antiferromagnetic nanoparticles \cite{book}.
These are composite objects with total spin $S$ ranging from $S=10$ for Fe-8 and
Mn-12 molecular clusters to $S$ of a few thousand in nanoparticles. Experimental study of the magnetization reversal in individual nanoparticles gives evidence that very small particles are uniformly magnetized \cite{Wernsdorfer1}. At low enough temperature they, like molecular magnets, possess a large fixed-length spin formed by the strong exchange interaction. 

Quantum tunneling between the $|\uparrow>$ and $|\downarrow>$ states of spin-10 molecular
nanomagnets has been unambiguously established in experiment \cite{Friedman,WS}. Early
measurements of low temperature magnetic relaxation \cite{book} and more recent measurements of individual nanoparticles \cite{Wernsdorfer2} also provided evidence
of quantum tunneling of spin. There is experimental evidence of quantum coherent oscillations between the $|\uparrow>$ and $|\downarrow>$ states in nanoparticles
\cite{Awschalom} and molecular magnets \cite{Barco}. 

On one hand, individual nanoparticles and high-spin molecules, because of their
large magnetic moments, must be easier to operate as qubits than individual electron 
and nuclear spins. In fact, the techniques of measuring individual particles of $S{\sim}10^{2}-10^{3}$ already exist \cite{Wernsdorfer3}. On the other hand, as 
the size of the system increases, its interaction with the dissipative environment 
also increases and it is not obvious whether the decoherence in large spin systems can be 
made low enough to allow their application as qubits. To achieve this goal the 
decoherence rate ${\gamma}$ must be made small compared to the frequency ${\Delta}/{\hbar}$ of coherent oscillations between the $|\uparrow>$ and $|\downarrow>$ states. For a large spin system it requires effort for two reasons. Firstly, the interaction with the environment is proportional to the size of the system. Secondly, $\log({\Delta})$ scales linearly with $-S$ so that, except for some special cases, ${\Delta}$ becomes immeasurably small for $S{\geq}30$. 

In this article we will make suggestions on how to obtain large ${\Delta}$ for 
a relatively large spin and how to achieve the condition ${\gamma}<<{\Delta}/{\hbar}$. 
It is our believe that this is doable in spin systems, in part due to their special 
properties with respect to time reversal. Spin Hamiltonians and tunneling rates will
be considered in Section 2. Mechanisms of decoherence will be discussed in Section 3. 

\section{Coherence}
\subsection{Ferromagnetic clusters}
The basic Hamiltonian for a ferromagnetic cluster of spin $S$ that exhibits
quantum coherence is
\begin{equation}
H=-AS_{z}^{2} + V\;\;,
\end{equation}
where $A$ is a positive constant and $V$ is an operator that does not commute
with $S_{z}$ but is invariant under the transformation $S_{z}{\rightarrow}-S_{z}$. 
The first term in Eq.(1) is typically produced by a crystal field or by the 
shape anisotropy of the cluster. At $V=0$ the Hamiltonian (1) has a
double degenerate ground state corresponding to two opposite orientations, 
$|\uparrow>$ and $|\downarrow>$, of ${\bf S}$ along the Z-axis. In terms of
the magnetic quantum number, $S_{z}|m>=m|m>$, these ground states are 
$|S>$ and $|-S>$. Note that this degeneracy is independent of any geometry 
of the problem including the shape of the magnetic cluster and is solely due
to the odd symmetry of ${\bf S}$ with respect to time reversal.

Our case of interest will be an integer large $S$. In that case the problem is
almost classical so that even a large perturbation $V{\neq}0$ generates only
a small probability of
tunneling between the $|\uparrow>$ and $|\downarrow>$ states. The degeneracy of 
the ground state is then removed and the new ground state can be approximated by
\begin{equation}
|0>=\frac{1}{\sqrt{2}}(|\uparrow>+|\downarrow>)\;\;.
\end{equation}
It is separated from the first excited state,
\begin{equation}
|1>=\frac{1}{\sqrt{2}}(|\uparrow>-|\downarrow>)\;\;,
\end{equation}
by the energy gap ${\Delta}$ which is determined by the strength and the nature
of $V$ and is small compared to the scale $A$ of the energy levels of $-AS_{z}^{2}$. 
The probability of finding ${\bf S}$ looking up (down) then oscillates in time
according $\cos({\Delta}{\cdot}t/{\hbar})$, which is quantum coherence. 

One of the possible forms of $V$ is $V=-g{\mu}_{B}H_{x}S_{x}$ induced by the 
external field applied along the X-axis. If the field is small, the tunneling splitting
can be obtained by the perturbation theory \cite{Garanin}
\begin{equation}
{\Delta}=\frac{4AS}{(2S-1)!}\left(\frac{g{\mu}_{B}|H_{x}|}{2A}\right)^{2S}\;\;.
\end{equation}
>From practical point of view this case is not very promising since the
inevitably present weak misorientation of the field, resulting in $H_{z}{\neq}0$, 
will destroy the coherence.

A more promising case, which to a good approximation corresponds to the Fe-8
spin-10 molecular nanomagnet, is $V=BS_{x}^{2}$. At $|B|<<A$ the perturbation
theory gives \cite{Garanin}
\begin{equation}
{\Delta}=8A\frac{(2S)!}{[(S-1)!]^{2}}\left(\frac{|B|}{16A}\right)^{S}\;\;.
\end{equation}
In Fe-8 ${\Delta}/{\hbar}$ is of order $10^{4}$s$^{-1}$. 

For large $S$ and
arbitrary $B$ the tunneling splitting has been computed by the instanton method 
\cite{Schilling,CG} and by mapping the spin problem onto a particle problem
\cite{Hemmen,Zaslavsky}
\begin{equation}
{\Delta}=16{\pi}^{-1/2}S^{3/2}|B|\left[\frac{A}{|B|}
\left(1+\frac{A}{|B|}\right)\right]^{3/4}
\left[\left(1+\frac{A}{|B|}\right)^{1/2}+
\left(\frac{A}{|B|}\right)^{1/2}\right]^{-(2S+1)}\;\;.
\end{equation}
>From practical point of view, in ferromagnetic nanoparticles with $S>>1$ 
the case of interest is $|B|>>A$. In that case Eq.(6) gives
\begin{equation}
{\Delta}=16{\pi}^{-1/2}S^{3/2}A^{3/4}|B|^{1/4}\exp\left[-2S\left(\frac{A}{|B|}
\right)^{1/2}\right]\;\;\;,
\end{equation}
and one can see that the effect of large $S$ in the exponent is suppressed by a
a small factor $(A/|B|)^{1/2}$. There are two ways to achieve this suppression and 
to increase ${\Delta}$. The first is to use magnetic clusters with 
very strong easy plane anisotropy and
relatively weak easy axis anisotropy in that plane. Particles of Tb and Dy may
satisfy this condition. The second way is to place the particle above the surface
of a superconductor \cite{CF}. In that case the magnetic dipole interaction of 
${\bf S}$ with its mirror image inside the superconductor effectively reduces
the uniaxial anisotropy $A$. The particle and the superconductor should be selected such that the magnetic field induced by the particle,
$H\,{\sim}\,4{\pi}{\mu}_{B}S$, does not exceed the first critical field of the 
superconductor, $H_{c1}$. Manipulating the distance between the particle
and the superconductor, one can achieve the condition $A<<|B|$. This can be done
by, e.g., controlling the distance with a ferroelectric buffer in the electric field.
Note that in the absence of the external magnetic field the odd symmetry of 
${\bf S}$ with respect to time reversal preserves the coherence of such a setup
independently of the shape of the superconducting surface and electric fields
in the problem. 

The tunneling of ${\bf S}$ also can be induced by its hyperfine interaction with 
nuclear spins, $V=B{\bf S}{\cdot}{\bf I}$, where ${\bf I}={\sum}{\bf I}_{i}$ is
the total nuclear spin of the cluster obtained by summing over spins of
individual nuclei. This problem is rather involved. It has been studied in Ref.\cite{GCS} and is relevant to tunneling in Mn-12. The total Hamiltonian conserves the magnitude
of $I$ and the Z-projection of ${\bf S}+{\bf I}$. 
In the millikelvin range nuclear spins must 
order, developing $I_{max}={\sum}|I_{i}|$. It is easy to see that the problem is the 
one of quantum coherence only if $I_{max}=S$. In that case the classical 
ground states correspond to ${\bf S}$ and ${\bf I}$ of equal length looking opposite
to each other along the Z-axis, $|S>|-I_{max}>$ and $|-S>|I_{max}>$. Tunneling removes
the degeneracy of the ground state. The corresponding splitting can be obtained by
the perturbation theory for $B<<A$:
\begin{equation}
{\Delta}=8(A+B)S^{2}\left[\frac{B}{2(A+B)}\right]^{2S}\;\;.
\end{equation}
In Mn-12 $S=10$ while $I_{max}=30$, so that the coherence of the above type is
impossible. It is not out of the question, however, that chemists will produce 
a molecular cluster with $S=I$ in the future.

\subsection{Antiferromagnetic clusters}

Tunneling in antiferromagnetic clusters \cite{BC} turns out to be much stronger than in
ferromagnetic clusters, making them promising candidates for quantum coherence.
Consider an anisotropic antiferromagnetic cluster with two compensated 
sublattices of spin ${\bf S}_{1}$ and ${\bf S}_{2}$, described by the Hamiltonian
\begin{equation}
\label{eq:AFM}
H= -A(S_{1z}^{2}+S_{2z}^{2})+B{\bf S}_{1}{\cdot}{\bf S}_{2}
\end{equation}
with positive $A$ and $B$ satisfying $A<<B$. 

Let us show that this model can be mapped onto the model with strong transverse anisotropy. The Lagrangian corresponding to Eq.(\ref{eq:AFM}) is \cite{book}
\begin{equation}
L=S({\dot{\phi}}_{1}\cos{\theta}_{1}+{\dot{\phi}}_{2}\cos{\theta}_{2})
-S({\dot{\phi}}_{1}+{\dot{\phi}}_{2})+AS^{2}({\cos}^{2}{\theta}_{1}+
{\cos}^{2}{\theta}_{2})-BS^{2}{\cos}{\theta}_{1}{\cos}{\theta}_{2}-
BS^{2}{\sin}{\theta}_{1}{\sin}{\theta}_{2}{\cos}({\phi}_{1}-{\phi}_{2})\;\;,
\end{equation}
where ${\phi}_{1}$, ${\theta}_{1}$, ${\phi}_{2}$, ${\theta}_{2}$ are spherical
coordinates of vectors ${\bf S}_{1}$ and ${\bf S}_{2}$ of fixed length $S$. 
The path integral over these angles is dominated by 
${\cos}{\theta}_{1}=-{\cos}{\theta}_{2}\,{\equiv}\,{\cos}{\theta}$, 
${\sin}{\theta}_{1}={\sin}{\theta}_{2}\,{\equiv}\,{\sin}{\theta}$. Introducing
${\phi}={\phi}_{1}-{\phi}_{2}$, one obtains, up to a phase term, the effective 
Lagrangian
\begin{equation}
L_{eff}=S{\dot{\phi}}{\cos}{\theta}+(2A+B)S^{2}{\cos}^{2}{\theta}-
BS^{2}{\sin}^{2}{\theta}{\cos}{\phi}\;\;.
\end{equation}
With the notations ${\phi}/2={\Phi}$ and $2S={\sigma}$, it can be transformed into
\begin{equation}
L_{eff}={\sigma}{\dot{\Phi}}{\cos{\theta}}+\frac{A}{2}{\sigma}^{2}{\cos}^{2}{\theta}
-\frac{B}{2}{\sigma}^{2}{\sin}^{2}{\theta}{\cos}^{2}{\Phi}\;\;,
\end{equation}
which is equivalent to the Hamiltonian
\begin{equation}
H=-\frac{A}{2}{\sigma}_{z}^{2} + \frac{B}{2}{\sigma}_{x}^{2}\;\;.
\end{equation}
One can then use the known result, Eq.(7), to obtain the tunneling splitting,
\begin{equation}
{\Delta}=32(2{\pi})^{-1/2}S^{3/2}A^{3/4}B^{1/4}\exp\left[-4S\left(
\frac{A}{B}\right)^{1/2}\right]\;\;.
\end{equation}
Here $B$ is the exchange constant which is typically $10^{4}-10^{6}$ the
anisotropy constant $A$. Consequently, antiferromagnetic particles consisting
of a few thousand magnetic atoms can exhibit a significant tunneling rate between the
$|{\uparrow}{\downarrow}>$ and $|{\downarrow}{\uparrow}>$ states. 

One should notice, that, in order to manipulate these states by the magnetic field, 
some magnetic non-compensation of the sublattices is needed. In this case tunneling is
possible only due to the presence in the Hamiltonian of the transverse field or the transverse anisotropy, e.g., $b(S_{1x}^{2}+S_{2x}^{2})$. Let the non-compensated 
spin be $s$. It has been demonstrated
\cite{MMM} that the antiferromagnetic tunneling with 
${\Delta}\,{\propto}\,\exp[-4S(A/B)^{1/2}]$ holds upto $s\,{\sim}\,(b/B)^{1/2}S<<S$. 
At greater $s$ it switches to the ferromagnetic tunneling with
${\Delta}\,{\propto}\,\exp[-2s(A/b)^{1/2}]$. Strongly non-compensated ferrimagnetic
clusters, like, e.g., Mn-12, are always in the ferromagnetic tunneling regime, while weakly non-compensated ferritin particles \cite{Awschalom,book} can be in the antiferromagnetic tunneling regime.

\section{Decoherence}

A magnetic cluster of the type described above will always be imbedded in a non-magnetic solid dissipative environment. The potential for low decoherence arises from a number of reasons. The first of them is that strong electrostatic interactions are involved,
through exchange couplings, only in the formation of the single spin ${\bf S}$ of the
cluster. All other interactions of ${\bf S}$ have relativistic smallness of 
order $(v/c)^{2}$ to a some power. Due to this fact the ferromagnetic resonance in some materials has a quality factor of one million. The second reason is selection rules for spins due to the time reversal symmetry discussed below.

In quantum computation one is interested in creating an arbitrary superposition of
the $|\uparrow>$ and $|\downarrow>$ states,
\begin{equation}
|{\Psi}> = C_{1}|{\uparrow}> + C_{2}|{\downarrow}>\;\;.
\end{equation}
Using equations (2) and (3) this state can be re-written in terms of $|0>$ and $|1>$:
\begin{equation}
|{\Psi}> = C'_{1}|0>+C'_{2}|1>\;\;,
\end{equation}
where $C'_{1}=(C_{1}+C_{2})/\sqrt{2}$ and $C'_{2}=(C_{1}-C_{2})/\sqrt{2}$.
It is clear, therefore, that the spontaneous decay of the excited state $|1>$ into
the ground state $|0>$, accompanied by the emission of the energy quantum 
${\Delta}$, should be a major concern for preserving quantum coherence. 

Note that, in principle, there may be decohering processes involving other excited 
levels of the spin Hamiltonian. For large spin, due to a small tunneling rate, such levels are separated from $|0>$ and $|1>$ by the energy gap (say $A$) that is large compared to ${\Delta}$. An example of such a process would be an Orbach two-phonon process 
corresponding to the transition $|1>\,{\rightarrow}\;|A>$ caused by the absorption of  a phonon, followed by the spontaneous decay $|A>\,{\rightarrow}\;|0>$ with the emission 
of a phonon. Also, there may be processes involving the excited states of the environment.
An example of such a process would be a two-phonon Raman process that corresponds to
the emission and absorption of two real phonons satisfying 
${\hbar}({\omega}_{1}-{\omega}_{2})={\Delta}$. All such processes are strongly temperature-dependent. Their strength is measured by $\exp(-A/k_{B}T)$ or by 
some high power of $T/{\Theta}$ where ${\Theta}{\sim}10^{2}K$ is the Debye temperature
\cite{AB}. In the millikelvin range the rate of such processes is negligible. 
Consequently, they are of little concern for the decoherence. On the contrary, processes of the spontaneous decay $|1>\,{\rightarrow}\;|0>$ can exist even at T=0. One should
therefore concentrate on such processes.

As follows from the previous section, cases of interest for quantum coherence are 
described by Hamiltonians which contain even powers of spin operators. Due to
the time reversal symmetry, this will always be the case in the absence of the external magnetic field. Consequently, the states $|0>$ and $|1>$ most generally can be 
written as
\begin{eqnarray}
|0> & = & \sum_{m=-S}^{S}{\alpha}_{m}|m> \\ \nonumber
|1> & = & \sum_{m=-S}^{S}{\beta}_{m}|m> \;\;,
\end{eqnarray}
where ${\alpha}_{m}={\alpha}_{-m}$ and ${\beta}_{m}=-{\beta}_{-m}$. 
These equations are exact, as compared to the approximate equations (2) and (3). They simply reflect the fact that $|m>$ is a complete set of vectors in the Gilbert space of the spin Hamiltonian and that $|0>$ and $|1>$ have different symmetry with respect to 
time reversal. 

Let $K$ be the antilinear antiunitary operator of time reversal. The spin operator is odd with respect to time reversal, $K{\bf S}K^{\dagger}=-{\bf S}$. On the contrary, the spin Hamiltonian that only contains even powers of components of ${\bf S}$ is even with 
respect to time reversal, $KHK^{\dagger}=H$. Consider now a decohering operator $D$.
$D$ can be due to the interaction of ${\bf S}$ with phonons, electromagnetic fields, nuclear spins, etc. These interactions have different symmetry with respect to time
reversal. Let $D_{o}$ and $D_{e}$ be time-odd and time-even operators respectively. 
Since the state $|O>$ is even with respect to time reversal and $|1>$ is odd, the following general statement is true
\begin{equation}
<0|D_{e}|1>=0 \;\;\;.
\end{equation} 

The spin-phonon interaction is of the form 
\begin{equation}
H_{sp}=a_{iklm}S_{i}S_{k}\frac{{\partial}u_{l}}{{\partial}r_{m}} + h.c. \;\;,
\end{equation}
where ${\bf u}$ is the lattice displacement and $a_{iklm}$ is a tensor reflecting the symmetry of the lattice. In terms of the operators of creation and annihilation of phonons
\begin{equation}
{\bf u}=\frac{i}{(2MN)^{1/2}}\sum_{{\bf k},{\lambda}}\frac{{\bf e}_{{\bf k},{\lambda}}
e^{i{\bf k}{\cdot}{\bf r}}}{({\omega}_{k{\lambda}})^{1/2}}
(a_{k{\lambda}}-a^{\dagger}_{k{\lambda}})\;\;,
\end{equation}
where $M$ is the unit cell mass, $N$ is the number of cells in the lattice,
${\bf e}_{{\bf k},{\lambda}}$ is the phonon polarization vector, ${\lambda}=t,t,l$,
and ${\omega}_{k{\lambda}}=v_{\lambda}k$ is the phonon frequency ($v_{\lambda}$
being the speed of sound). The spin-phonon
interaction given by Eq.(19) describes transitions between different spin states
accompanied by the emission and absorption of phonons. For instance, in the absence of
tunneling the excited state $|S-1>$ of the Hamiltonian $H=-AS_{z}^{2}$, corresponding to the ferromagnetic resonance, can relax to the ground state $|S>$ at $T=0$ by spontaneously emitting a phonon at a rate \cite{AB}
\begin{equation}
{\gamma}=C\frac{A^{2}S^{2}{\Delta}^{3}}{{\hbar}^{4}{\rho}v^{5}}\;\;,
\end{equation}
where ${\rho}$ is the mass density of the lattice and $C$ is a constant of
order unity. For, e.g., $A\,{\sim}\,1K$, $S\,{\sim}\,10^{3}$, 
$({\Delta}/{\hbar})\,{\sim}\,10^{9}s^{-1}$, ${\rho}\,{\sim}\,1 g/cm^{3}$, and 
$v\,{\sim}\,10^{5}cm/s$, Eq.(21) gives ${\gamma}\,{\sim}\,10^{3}s^{-1}$. 

We should now notice that $KH_{sp}K^{\dagger}=H_{sp}$, that is, the spin-phonon operator 
is even with respect to time reversal. Consequently, $<0|H_{sp}|1>=0$. This also can be 
checked by the direct calculation of matrix elements of $H_{sp}$ using 
expressions (17). Thus, $|1>$ cannot spontaneously decay into $|0>$ with an emission
of a phonon and we conclude that in a millikelvin range the spontaneous emission
of phonons cannot decohere the coherent superposition of the $|\uparrow>$ and $|\downarrow>$ spin states. This
is a strong statement which requires some clarification. Indeed, the spin-phonon interaction originates from the spin-orbit coupling of the form 
$H_{so}\,{\propto}\,{\bf L}{\cdot}{\bf S}$ where ${\bf L}$ is the orbital momentum. Its density can be presented as ${\rho}{\epsilon}_{ikl}r_{k}{\dot{u}}_{l}$. This operator
seems to have non-zero matrix elements between $|0>$ and $|1>$ given by equations (17).
However, this is only because the formulation of the spin tunneling problem presented
above does not account for the conservation of the total angular momentum. In fact,
the coherence is possible only if ${\bf L}+{\bf S}=0$ \cite{ECPRL}. Consequently,
as ${\bf S}$ tunnels between $|\uparrow>$ and $|\downarrow>$, so does the mechanical
angular momentum of the body. This is an analogue of
the M\"{o}ssbauer effect for spin tunneling. The energy associated with the mechanical
rotation is ${\hbar}^{2}S^{2}/2I_{in}$ where $I_{in}$ is the moment of inertia of the solid matrix containing
the magnetic cluster. This energy should not exceed ${\Delta}$, otherwise it would
be energetically favorable for ${\bf S}$ to localize in the $|\uparrow>$ or
$|\downarrow>$ state. Since $I_{in}$ scales as the fifth power of the size of the matrix,
it is easy to see that such localization may occur in a free particle of size less 
then 10 nm,
while for bigger systems the conservation of the total angular momentum is 
rather formal than practical question. With the account of the momentum conservation,
the ground state and the first excited state are 
$\frac{1}{\sqrt{2}}(|S>|-L>\,{\pm}\,|-S>|L>)$, where the absolute values of ${\bf S}$
and ${\bf L}$ are equal. Now, again, the $|0>$ state is even with respect to time reversal while the $|1>$ state is odd. The spin-orbit operator is even, 
$KH_{so}K^{\dagger}=H_{so}$, because ${\bf L}$ is odd, $K{\bf L}K^{\dagger}=-{\bf L}$.
Thus, as before, $<0|H_{so}|1>=0$. 

As has been shown in the previous section (see also Refs. 21,22), nuclear 
spins always destroy the coherence at $H=0$ unless tunneling is induced by the hyperfine interaction with $I=S$. Except for that exotic possibility, nuclear spins must be always eliminated from magnetic qubits by the isotopic purification. 
Similarly, the presence of free non-superconducting electrons in the sample will decohere tunneling through the spin scattering of electrons, 
$D_{e}\,{\propto}\,{\bf s}{\cdot}{\bf S}$.
Although, this operator is time-even, free electrons incidentally passing through the magnetic cluster, perturb $|0>$ and $|1>$, breaking their properties with respect to
time reversal. Thus, strongly insulating materials should be chosen for magnetic qubits. The effect of incidental phonons due to, e.g., relaxation of elastic stresses in the
matrix (the $1/f$ noise), should be similar to the effect of incidental electrons in
perturbing $|{\Psi}>$. Thus, the perfection of the lattice should be given a serious
thought when manufacturing magnetic or any other qubits. One should then worry about the decohering effect of spin interactions which are odd with respect to time reversal. 
These are Zeeman terms due to magnetic fields. 
For example, $D_{o}=-g{\mu}_{B}S_{z}H_{z}(t)$ has a non-zero matrix
element between $|0>$ and $|1>$. The effort should be made, therefore, to shield the
magnetic qubit from the magnetic fields during the process of quantum computation. This can be done by placing the magnetic cluster inside a nanoscopic superconducting ring. 
Such a ring may be used to control and measure the states of the qubit. Connecting such
rings by superconducting lines may be the way to make a miltiqubit system. 

\section*{Acknowledgements}

I am grateful to Joe Birman, Lev Bulaevski, and Dima Garanin for discussions.
This work has been supported by the U.S. National Science Foundation Grant 
No.\ DMR-9978882.

% =================================================================== 
%  References

%
\end{document}